\documentclass[
  twocolumn,
  superscriptaddress,
  aps,
  prc,
  nofootinbib
]{revtex4-2}
\usepackage{amssymb}
\usepackage[centertags]{amsmath}
\usepackage{graphicx,subfigure}
\usepackage{txfonts}
\usepackage{epsfig}
\usepackage{bm}
\usepackage{color}
\usepackage{graphicx,graphics}
\usepackage{multirow}
\usepackage{float}
\usepackage{appendix}
\usepackage{changes}

\allowdisplaybreaks[2]

\usepackage{graphicx}   
\usepackage{float} 
\usepackage{xspace}     
\usepackage{url}
\usepackage[mathlines]{lineno}
\usepackage{bm}
\usepackage{amsmath}
\usepackage{xcolor}
\usepackage[driverfallback=dvipdfm]{hyperref}
\usepackage{subfigure}
\hypersetup{
	colorlinks = true,
	citecolor= blue,
	linkbordercolor = cyan
}

\def \be {\begin{equation}}
\def \ee {\end{equation}} 
\def \bea {\begin{eqnarray}}
\def \eea {\end{eqnarray}}

\def \snn {$\sqrt{s_{_{\rm NN}}}$\xspace} 
\usepackage{color}
\definecolor{orange}{cmyk}{0.,0.353,1.,0.}    

\usepackage{lineno}
\newcommand*\patchAmsMathEnvironmentForLineno[1]{
  \expandafter\let\csname old#1\expandafter\endcsname\csname #1\endcsname
  \expandafter\let\csname oldend#1\expandafter\endcsname\csname end#1\endcsname
  \renewenvironment{#1}
     {\linenomath\csname old#1\endcsname}
     {\csname oldend#1\endcsname\endlinenomath}}
\newcommand*\patchBothAmsMathEnvironmentsForLineno[1]{
  \patchAmsMathEnvironmentForLineno{#1}
  \patchAmsMathEnvironmentForLineno{#1*}}
\AtBeginDocument{
\patchBothAmsMathEnvironmentsForLineno{equation}
\patchBothAmsMathEnvironmentsForLineno{align}
\patchBothAmsMathEnvironmentsForLineno{flalign}
\patchBothAmsMathEnvironmentsForLineno{alignat}
\patchBothAmsMathEnvironmentsForLineno{gather}
\patchBothAmsMathEnvironmentsForLineno{multline}
}


\AtBeginDocument{%
  \renewcommand{\textsc}[1]{\textit{#1}}%
}

\usepackage[T1]{fontenc}  

\begin{document}

\title{Differential measurement of global polarization of $\Lambda$, $\Xi$, and $\Omega$ hyperons in Au+Au collisions at $\sqrt{s_{\mathrm{NN}}}=7.7$--$27~\mathrm{GeV}$ from the RHIC Beam Energy Scan-II}

\author{The STAR Collaboration}

\begin{abstract}
We present high-precision differential measurements of the global polarization of $\Lambda$, $\Xi$, and $\Omega$ hyperons in Au+Au collisions at $\sqrt{s_{\mathrm{NN}}}=7.7$, $9.2$, $11.5$, $14.6$, $17.3$, $19.6$, and $27~\mathrm{GeV}$ from the RHIC Beam Energy Scan Phase II (BES-II) using the STAR detector. The results show a clear centrality dependence, with polarization increasing toward more peripheral collisions. 
Within the present experimental precision, no significant dependence on transverse momentum is observed for $\Lambda$ and $\Xi$, and no significant rapidity dependence is observed for the $\Xi$.
The polarization of $\Xi$ hyperons is comparable at the current precision to that of $\Lambda$ hyperons, consistent with a picture in which all hyperons acquire polarization from the same global angular momentum source, with differences arising from quark content and hadronization dynamics. 
These high-precision, differential measurements across centrality and $p_T$ over the measured energy ranges provide a unique test of spin--vorticity coupling and sensitivity to spin--spin correlations among hyperons with different quark content (uds, dss, and sss). The results are consistent, within current precision and under the stated spin-transfer assumptions, with a common vorticity-driven polarization mechanism across strange and multistrange hyperons. Consequently, they constrain the interplay of vorticity, magnetic fields, and quark-flavor dynamics in the produced medium.
\end{abstract}

\maketitle

\section{Introduction}
Spin is a fundamental quantum degree of freedom that plays an essential role across nuclear and particle physics. It governs the nuclear shell structure, contributes to the spin decomposition of the nucleon, and provides sensitivity to the microscopic properties of strongly interacting matter~\cite{STAR:2005gfr, PHENIX:2004vcz,BRAHMS:2004adc,PHOBOS:2004zne,ALICE:2022wpn}. In relativistic heavy-ion collisions, the study of spin degrees of freedom has gained increasing attention because such collisions generate systems characterized by extremely large vorticity, electromagnetic fields, and baryon densities. These conditions offer a unique opportunity to investigate the coupling between spin and the macroscopic properties of the medium created in the collision.

Through spin--orbit interactions in the hot and rapidly rotating medium produced in non-central heavy-ion collisions, the system's large orbital angular momentum can induce a polarization of spin-bearing hadrons along its direction, referred to as global polarization~\cite{1, Gao:2007bc,2,STAR:2007ccu,STAR:2017ckg,STAR:2022fan}. The STAR Collaboration provided the first experimental evidence for global spin polarization via the observation of nonzero $\Lambda$ and $\bar{\Lambda}$ polarization in Au+Au collisions at RHIC energies~\cite{STAR:2017ckg}. These measurements demonstrated that the created medium exhibits exceptionally large vorticity, far beyond that of any known macroscopic fluid. Results from the first phase of the RHIC Beam Energy Scan (BES-I) further established systematic features of this phenomenon, including a monotonic decrease of global polarization with increasing collision energy. Both hydrodynamic and transport calculations reproduce this trend, consistent with expectations of reduced baryon stopping at higher $\sqrt{s_{\mathrm{NN}}}$~\cite{Becattini:2024uha,Karpenko:2016jyx,Vitiuk:2019rfv,Voronyuk:2023vyu,Yi:2024kwu,Sheng:2024pbw,Fang:2023bbw,Becattini:2020ngo,Niida:2024ntm,Chen:2024aom, Wu:2022mkr,Chen:2026gka}.
BES-I results also hinted at a possible difference between $\Lambda$ and $\bar{\Lambda}$ polarizations. Although not statistically significant, the largest effect was seen at $\sqrt{s_{\mathrm{NN}}}=7.7$ GeV with a significance of about two standard deviations. This stimulated considerable theoretical discussion regarding potential contributions from the strong magnetic field generated early in the collision~\cite{Guo:2019joy,Becattini:2016gvu}. 
Later high-statistics measurements by STAR at 19.6 and 27 GeV~\cite{STAR:2023nvo} found no significant splitting, indicating that the possible effect does not manifest with a large magnitude at higher energies. 
Additional sensitivity to the underlying spin dynamics comes from multistrange hyperons such as $\Xi$ and $\Omega^{-}$, whose small hadronic interaction cross sections and different quark content make them valuable probes of early-stage vorticity and flavor-dependent effects~\cite{Li:2021zwq}. STAR results from Au+Au and isobar (Ru+Ru and Zr+Zr) collisions at 200 GeV showed that the polarizations of $\Lambda$, $\bar{\Lambda}$, $\Xi^{-}+\bar{\Xi}^{+}$, and $\Omega^{-}+\bar{\Omega}^{+}$ hyperons are consistent within uncertainties, supporting the picture of a collective and system-wide spin polarization mechanism~\cite{STAR:2020xbm, STAR:2018gyt, STAR:2025dgs}.

The second phase of the RHIC Beam Energy Scan (BES-II), spanning $\sqrt{s_{\mathrm{NN}}}=7.7$--$27~\mathrm{GeV}$, benefits from greatly increased statistics and major detector upgrades, which together significantly improve the precision and kinematic reach of hyperon polarization measurements. 
Leveraging these enhanced capabilities, this analysis presents differential measurements of the global polarization of $\Lambda$, $\Xi$, and $\Omega$ hyperons in Au+Au collisions across the BES-II energy range. The collision-energy dependence of the hyperon polarizations and the rapidity dependence of the $\Lambda$ polarization from BES-II were reported in Ref.~\cite{shortpaper}. Here, we provide the full differential characterization, including the centrality, transverse-momentum, and rapidity dependencies, together with expanded systematic studies. These results constitute the most precise and comprehensive measurements of hyperon global polarization at intermediate collision energies to date. The expanded species coverage, improved statistical precision, and enhanced differential constraints enable more stringent tests of theoretical descriptions and provide new insight into the spin–vorticity coupling in relativistic heavy-ion collisions. 

The remainder of this paper is organized as follows: Section~\ref{sec:conservation} describes the detector subsystems, datasets, event selection and charged-particle track selection criteria, the reconstruction of hyperons, the methodology used to extract global polarization and the estimation of systematic uncertainties. Section~\ref{sec:results} presents the results as functions of collision centrality, transverse momentum and rapidity.
Section~\ref{sec:summary} summarizes the main findings.

\section{\label{sec:conservation}Experimental details }
\subsection{The STAR detector}

The STAR detector~\cite{STAR:2002eio} is a large-acceptance experiment designed to reconstruct the geometry and particle content of relativistic heavy-ion collisions. The primary subsystems used are the Time Projection Chamber (TPC)~\cite{Anderson:2003ur}, the Time-of-Flight (TOF)~\cite{Llope:2012zz} detector, the Event Plane Detector (EPD)~\cite{Adams:2019fpo}, and the Vertex Position Detectors (VPDs)~\cite{Llope:2014nva}. The TPC, located at the center of STAR and immersed in a 0.5~T solenoidal magnetic field, provides full azimuthal coverage and a pseudorapidity acceptance of $|\eta| < 1.0$. It serves as the primary tracking device, measuring charged-particle momenta from their curvature in the magnetic field and providing particle identification through specific ionization energy loss ($dE/dx$). During the BES-II program, the installation of the inner TPC (iTPC) upgrade~\cite{Shen:2018pkc} extended the pseudorapidity coverage to $|\eta| < 1.5$ and increased the number of readout rows, improving momentum resolution, tracking efficiency, and sensitivity to low-$p_{T}$ charged particles. 
For high-quality reconstruction, charged-particle tracks are required to have at least 15 TPC hits, with the number of hits used in the track fit being at least 52\% of the number of hits possible along the reconstructed track trajectory.
The TPC also determines the primary collision vertex position along and transverse to the beam axis.

Surrounding the TPC, the TOF detector augments particle identification through precise timing measurements. The system uses multigap resistive plate chamber technology arranged in 120 trays around the TPC, covering $|\eta| < 0.9$ with full azimuthal acceptance. Combined with start-time information from the VPDs, the TOF achieves a timing resolution of approximately 100~ps, extending particle identification beyond the momentum region where $dE/dx$ alone is effective, up to transverse momenta of $p_{T} \approx 3~\mathrm{GeV}/c$. 

Forward-rapidity detectors play a central role in determining the event geometry. The EPD consists of two scintillator wheels placed at $\pm 3.75~\mathrm{m}$ from the interaction point and covering $2.1 < |\eta| < 5.1$. Each wheel is segmented into radial and azimuthal tiles, enabling detailed measurements of forward particle multiplicity and azimuthal anisotropies. Compared with the earlier Beam--Beam Counters, the EPD provides substantially improved granularity and acceptance, yielding a significant enhancement in first-order event-plane resolution, especially at the lower collision energies of BES-II. 

\subsection{Data set}

The data were collected from Au+Au collisions at $\sqrt{s_{\mathrm{NN}}} = 7.7, 9.2, 11.5, 14.6, 17.3, 19.6,$ and $27~\mathrm{GeV}$ between 2018 and 2021 during the RHIC Beam Energy Scan-II.
Primary vertices are selected from the center of STAR with a transverse radial position of $v_{r} < 2.0~\mathrm{cm}$ and a longitudinal position of $|v_z| < 145~\mathrm{cm}$ for $\sqrt{s_{\mathrm{NN}}} = 7.7$--$19.6~\mathrm{GeV}$ and $|v_z| < 70~\mathrm{cm}$ for $\sqrt{s_{\mathrm{NN}}} = 27~\mathrm{GeV}$. 
Data are selected with the minimum-bias trigger, requiring hits in both VPDs,
located at $4.23 < |\eta| < 5.1$, and the Zero Degree Calorimeters (ZDCs)~\cite{Adler:2001fq} at $|\eta| > 6.3$. The out-of-time pile-up events, a beam-induced background where multiple collisions are recorded as one event, were rejected using a correlation between the total number of tracks in the TPC and the number of tracks matched with a hit in the TOF detector. The collision centrality is estimated by comparing the measured multiplicity distribution, obtained from tracks within $|\eta| < 0.5$, with Monte Carlo Glauber simulations~\cite{STAR:2012och}. Finally, detailed quality-assurance tests were performed to ensure that only events for which RHIC and the relevant detectors were performing adequately are included in this study. Details of the data set and the event vertex selection criteria, as well as the event statistics reflecting the allocated beam time and RHIC performance during each running period, are summarized in Table~\ref{tab:data_sets}. 

\begin{table}[!ht]
    \centering
    \begin{tabular}{cccccc}
        \hline
        $\sqrt{s_{\mathrm{NN}}}$ (GeV) & Year & $|v_z|$ (cm) &  Events~(M) & Track $|\eta|$ & Hyperon $|\eta|$  \\
        \hline
        7.7  & 2021 & $< 145$ & 110 & $<1.5$ & $<1.5$  \\
        9.2  & 2020 & $< 145$ & 204 & $<1.5$ & $<1.5$  \\
        11.5 & 2020 & $< 145$ & 269 & $<1.5$ & $<1.5$  \\
        14.6 & 2019 & $< 145$ & 413 & $<1.5$ & $<1.5$  \\
        17.3 & 2021 & $< 145$ & 275 & $<1.5$ & $<1.5$  \\
        19.6 & 2019 & $< 145$ & 668 & $<1.5$ & $<1.5$  \\
        27   & 2018 & $< 70$  & 521 & $<1.0$ & $<1.0$  \\
        \hline
    \end{tabular}
    \caption{Summary of datasets, vertex selection, number of analyzed events, track and hyperon acceptance in Au+Au collisions at different energies.}
    \label{tab:data_sets}
\end{table}

\subsection{Event plane reconstruction}
\label{Sec_resoolution}

In non-central heavy-ion collisions, the system's orbital angular momentum is oriented perpendicular to the reaction plane, defined by the beam direction and the impact-parameter vector $\vec{b}$ connecting the centers of the two colliding nuclei~\cite{Adams:2021yob}. While the true reaction plane cannot be directly measured, its azimuthal orientation can be estimated on an event-by-event basis using the ``event plane method'', which relies on the anisotropic emission of particles preferentially along the reaction plane.

The first-order event plane angle, $\Psi_1$, is reconstructed using the Event Plane Detector (EPD) and the ``Q-vector method''. When particles from the collision hit the detector, their spatial positions and energy depositions are recorded. Each hit point defines a vector $\vec{Q}_i$ connecting the hit position to the detector center, and the total Q-vector for the event, $\vec{Q}_n$, is obtained by summing over all $\vec{Q}_i$~\cite{Poskanzer:1998yz, Voloshin:2008dg}. 
The contribution of each EPD tile to $\Psi_1$ is weighted by its signal strength, which correlates with the multiplicity in that tile, and further weighted by the measured directed flow at the corresponding pseudorapidity~\cite{STAR:2023nvo}.

Due to detector limitations, the reconstructed event plane may deviate from the true reaction plane. This deviation is quantified by the first-order event-plane resolution, $\rm Res(\Psi_1) = \langle \cos(\Psi_1 - \Psi_{\rm RP}) \rangle$, where $\Psi_{\rm RP}$ denotes the reaction plane angle. For symmetric collision systems, $\rm Res(\Psi_1)$ can be determined from correlations between $\Psi_1$ measurements obtained from the EPD at forward (west) and backward (east) rapidities~\cite{Poskanzer:1998yz, Voloshin:2008dg}. Figure~1 shows $\rm Res(\Psi_1)$ as a function of collision centrality for $\sqrt{s_{\mathrm{NN}}} = 7.7, 9.2, 11.5, 14.6, 17.3, 19.6,$ and $27\ \mathrm{GeV}$. For mid-central collisions, the maximum value of $\rm Res(\Psi_1)$ decreases with increasing beam energy, ranging from approximately \textcolor{blue}{0.75} at 7.7 GeV to \textcolor{blue}{0.5} at 27 GeV. 
The EPD resolution is approximately 50\% higher than that obtained with the Beam--Beam Counters used in the BES-I analysis~\cite{STAR:2017ckg}. Since the statistical sensitivity of the resolution-corrected polarization measurement scales approximately with $1/\rm Res(\Psi_1)$ for a fixed event sample, the improved event-plane resolution provides a substantial gain in polarization sensitivity in BES-II.

\begin{figure}[H]
    \centering
    \vspace{-4mm}
    \hspace{-3mm}
    \includegraphics[width=0.75\linewidth,clip]{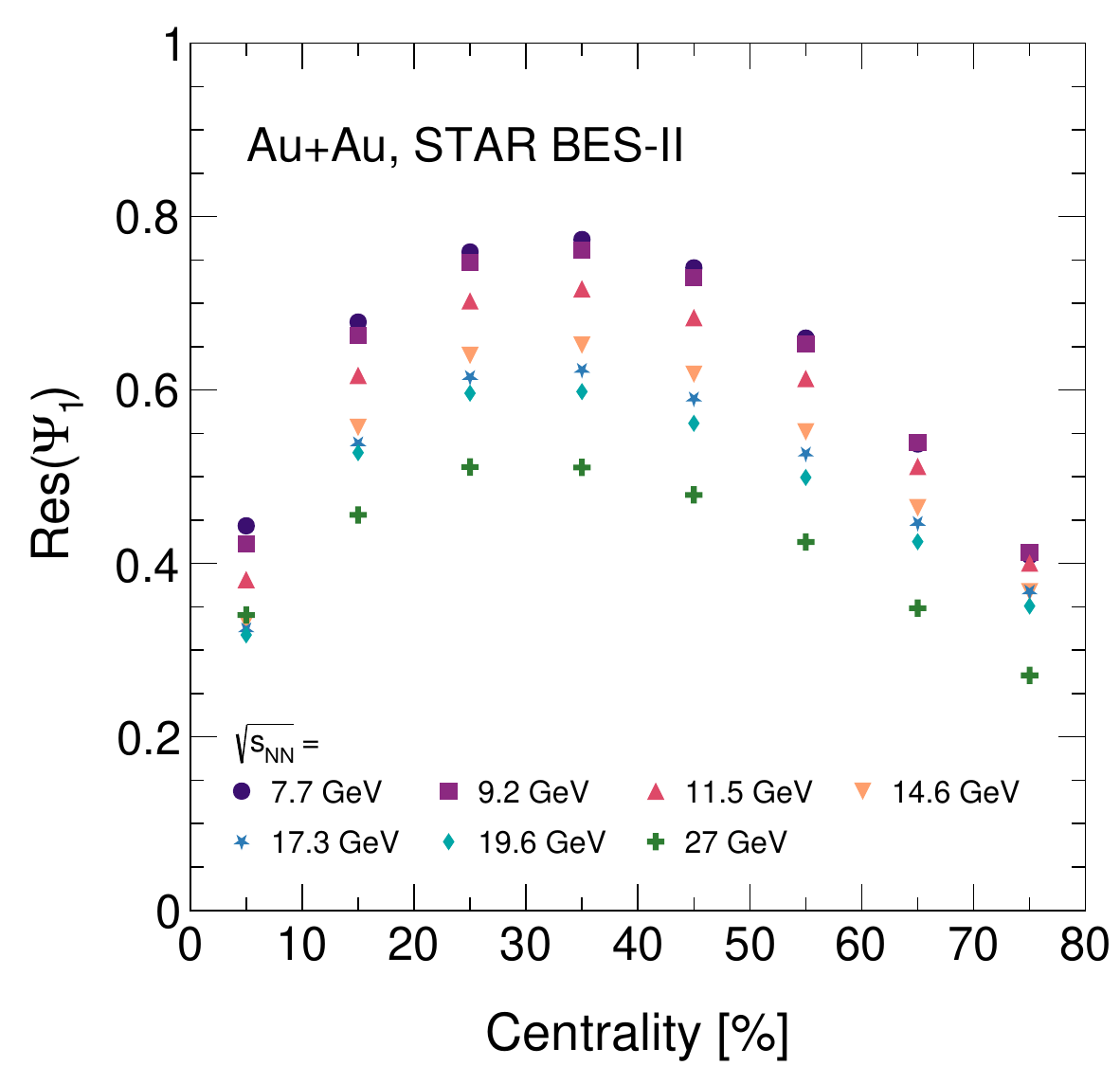}
    \vspace{-2mm}
    \caption{The first-order event-plane resolution, $\rm Res(\Psi_1)$, from the EPD as a function of collision centrality in Au+Au collisions at \snn = 7.7, 9.2, 11.5, 14.6, 17.3, 19.6, and 27 GeV.}
    \label{fig:EPDPsi1Resolution}
\end{figure}

\subsection{Hyperon reconstruction}

Hyperons are reconstructed using both the traditional topological method and the KFParticle algorithm~\cite{Gorbunov:2013yvt,Zyzak:2016exl,Kisel:2018nvd, STAR:2017ckg,STAR:2023nvo,STAR:2020xbm}, which performs Kalman-filter-based secondary vertex reconstruction. The species considered include $\Lambda$ and $\bar{\Lambda}$, $\Xi^{-}$ and $\bar{\Xi}^{+}$, and $\Omega^{-}$ and $\bar{\Omega}^{+}$. 
$\Lambda$ and $\bar{\Lambda}$ hyperons are neutral particles that decay weakly via $\Lambda \rightarrow p + \pi^{-}$ and $\bar{\Lambda} \rightarrow \bar{p} + \pi^{+}$ (branching ratio 63.9\%). Reconstruction proceeds through their daughter particles, which are identified using ionization energy loss in the TPC and mass-squared $m^{2}$ information from the TOF detectors. The secondary decay vertex (V$^0$) is determined from the distance of closest approach (DCA) of the daughter tracks. The decay length, defined along the reconstructed hyperon flight direction from the primary to the secondary vertex, and the angle between this flight direction and the mother-particle momentum [$\cos(A)$] are used for topological selections. The basic track selection criteria and optimized topological cuts used to maximize the signal-to-noise ratio for $\Lambda$ and $\bar{\Lambda}$ are summarized in Table~\ref{tab:TopolCutsSummary}.
For $\Lambda$ and $\bar{\Lambda}$ hyperons, the optimized topological reconstruction yields slightly higher signal significance than KFParticle, particularly in the extended pseudorapidity region $1.0<|\eta|<1.5$, and is therefore used as the default method. For the lower-yield $\Xi$ and $\Omega$ hyperons, KFParticle provides more efficient multivertex reconstruction and stronger background suppression and is adopted as the default. The $\Lambda$ polarization results obtained with the two reconstruction approaches are statistically consistent.

\begin{figure}[H]
    \centering
    \includegraphics[width=0.75\linewidth]{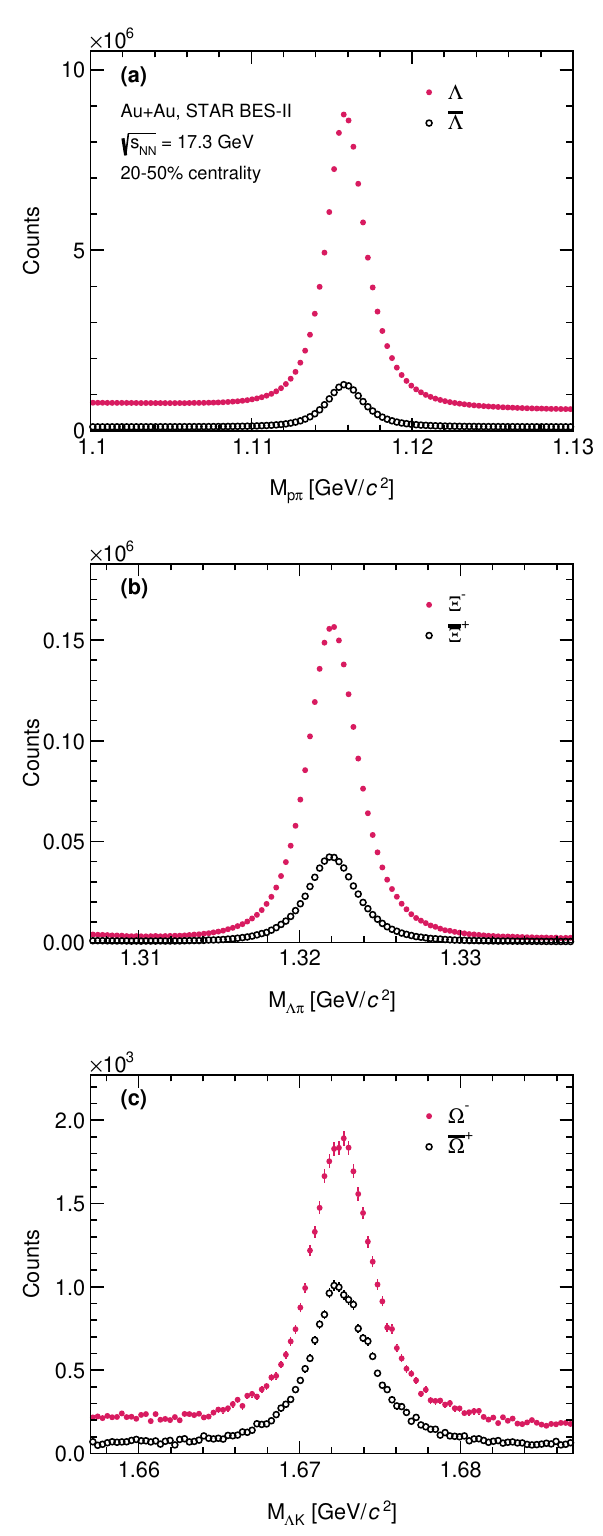}
    \caption{Invariant mass distributions for $\Lambda$($\bar{\Lambda}$) (a), $\Xi^{-}$($\bar{\Xi}^{+}$) (b), and $\Omega^{-}$($\bar{\Omega}^{+}$) (c) for 20--50\% centrality in Au+Au collisions at $\sqrt{s_{\mathrm{NN}}}=17.3~\mathrm{GeV}$.}
    \label{fig:LambdaMass}
\end{figure}

For the KFParticle reconstruction, global tracks are pre-selected with basic quality criteria and PID information from the TPC and TOF detectors. 
Identified particles are provided as input to KFParticle, which performs secondary vertex finding and topological reconstruction. Using these inputs, KFParticle reconstructs secondary vertices and extracts signals for $\Xi^{-}$, $\bar{\Xi}^{+}$, $\Omega^{-}$, and $\bar{\Omega}^{+}$. Invariant mass distributions are fitted to obtain signal yields and purities. KFParticle provides good purities for all centrality bins. 
Figure~\ref{fig:LambdaMass} shows representative invariant-mass distributions for $\Lambda$ and $\bar{\Lambda}$, $\Xi^{-}$ and $\bar{\Xi}^{+}$, and $\Omega^{-}$ and $\bar{\Omega}^{+}$ in 20--50\% central Au+Au collisions at $\sqrt{s_{\mathrm{NN}}}=17.3~\mathrm{GeV}$.

\begin{table}[H]
    \centering
    \begin{tabular}{|c|c|c|}
    \hline
        Variable & $\Lambda$ & $\bar{\Lambda}$ \\ \hline
        $p_T$ (GeV/$c$) & 0.5--6.0 & 0.5--6.0 \\
        $|\eta|$ & $<1.5$ & $<1.5$ \\
        $\cos(A)$ & $>0.99$ & $>0.99$ \\
        $p$ DCA (cm) & $>0.29$ & $>0.40$ \\
        $\pi$ DCA (cm) & $>0.68$ & $>0.90$ \\
        $p$--$\pi$ DCA (cm) & $<1.0$ & $<0.9$ \\
        $\Lambda$ Decay length (cm) & $>2.9$ & $>2.5$ \\
        $\Lambda$ DCA (cm) & $<0.9$ & $<1.0$ \\
    \hline
    \end{tabular}
    \caption{Optimized topological cuts for $\Lambda$ and $\bar{\Lambda}$. }
    \label{tab:TopolCutsSummary}
\end{table}

\subsection{\label{Polarization measurement}Polarization measurement}
\label{sec:PHextract}

The global polarization of hyperons, $P_H$, can be probed through their weak decays, as the daughter baryon is preferentially emitted along the hyperon spin direction with the opposite orientation for anti-hyperons. Therefore, $P_H$ can be extracted from the angular distribution of hyperon decay products relative to the system's orbital angular momentum~\cite{STAR:2007ccu}. In practice, detector limitations can restrict the angular coverage of emitted protons due to acceptance or efficiency effects. 
As in previous STAR analyses, the hyperon polarization is extracted as

\begin{equation}
\label{LambdaPHfunction}
    P_{\rm H} = \frac{8}{\pi\alpha_{\rm H}} \frac{1}{A_0} \frac{\langle\sin(\Psi_1-\phi_{\rm D}^{\ast})\rangle}{\rm Res(\Psi_1)},
\end{equation}
where $A_0$ is an acceptance correction term estimated following the approach of previous STAR analyses~\cite{STAR:2007ccu}, $\alpha_H$ is the hyperon decay parameter, $\phi_D^*$ denotes the azimuthal angle of the daughter baryon in the hyperon rest frame, and $\rm Res(\Psi_1)$ denotes the first-order event-plane resolution. For $\Lambda$ hyperons, $\alpha_\Lambda = -\alpha_{\bar{\Lambda}} = 0.732 \pm 0.014$~\cite{ParticleDataGroup:2020ssz}, and $\phi_D^*$ corresponds to the azimuthal angle of the daughter proton in the $\Lambda$ rest frame.

For $\Xi^{-}$ hyperons, $\phi_D^*$ is defined as the azimuthal angle of the daughter $\Lambda$ in the $\Xi^{-}$ rest frame, with $\alpha_{\Xi^{-}} = -\alpha_{\bar{\Xi}^{+}} = -0.401 \pm 0.010$~\cite{ParticleDataGroup:2020ssz}. The smaller decay parameter, together with the lower yield of $\Xi$ hyperons, leads to larger statistical uncertainties. Two methods are used to extract the $\Xi^{-}$ polarization: the direct method uses the angular distribution of the daughter $\Lambda$ in the parent $\Xi^{-}$ rest frame together with $\alpha_{\Xi^{-}}$; the indirect method first determines the daughter $\Lambda$ polarization from its proton decay distribution and then infers the parent $\Xi^{-}$ polarization via the spin-transfer factor $C_{\Xi\Lambda}$.
The relation between the parent and daughter polarizations is given by \(P_\Lambda^* = C_{\Xi\Lambda} \, P_\Xi^*\)~\cite{STAR:2020xbm, Lee:1957qs, E756:2000rge, HyperCP:2004not, ParticleDataGroup:2020ssz, Luk:1988as, Kim:1992az}. The spin-transfer factor for the decay \(\Xi^{-} \rightarrow \Lambda \pi^-\) is
\begin{equation}
    C_{\Xi\Lambda} = \frac{1}{3} (1 + 2 \gamma_\Xi),
\end{equation}
where \(\gamma_\Xi = 0.916\) is the measured parameter~\cite{HyperCP:2004not,STAR:2020xbm}, yielding \(C_{\Xi\Lambda} = 0.944\).

For $\Omega^{-}$ hyperons, the small decay parameter $\alpha_{\Omega}=0.0157\pm0.0021$~\cite{ParticleDataGroup:2020ssz} makes a direct determination of the parent polarization from the daughter-$\Lambda$ angular distribution impractical.
The reported $\Omega^{-}$ polarization is indirectly measured via the daughter $\Lambda$ polarization under an assumption for the spin-transfer factor,
\begin{equation}
    C_{\Omega\Lambda} = \frac{1}{5} (1 + 4 \gamma_\Omega).
\end{equation}
The time-reversal-violation parameter $\beta_\Omega$ is expected to be small.
Together with the normalization condition $\alpha_\Omega^2+\beta_\Omega^2+\gamma_\Omega^2=1$, this gives the approximate solutions $\gamma_\Omega\approx\pm1$, corresponding to $C_{\Omega\Lambda}\approx+1$ and $C_{\Omega\Lambda}\approx-0.6$, respectively~\cite{STAR:2020xbm}.
The reported $\Omega^{-}$ polarization assumes $C_{\Omega\Lambda}\approx+1$; the alternative solution would reverse the sign of the inferred parent polarization.

\begin{figure}[H]
    \centering
    \hspace{-2mm}
    \includegraphics[width=0.70\linewidth]{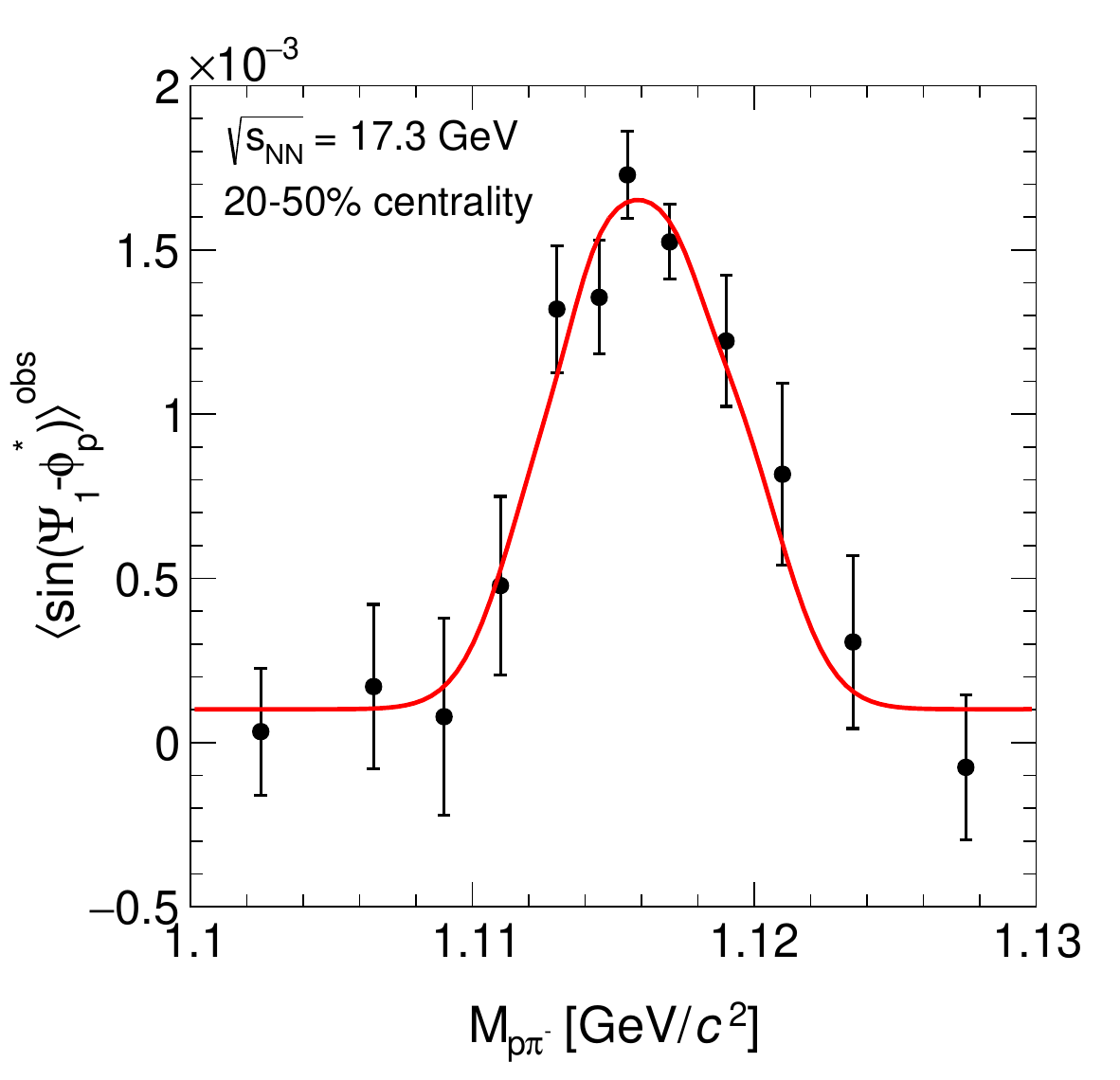}
    \vspace{-2mm}
    \caption{
    $\langle \sin(\Psi_1 - \phi_p^*) \rangle^{\rm obs}$ as a function of invariant mass for $\Lambda$ at 20--50\% centrality in Au+Au collisions at $\sqrt{s_{\mathrm{NN}}}=17.3~\mathrm{GeV}$. The IM method is used to extract the global polarization signal.}
    \label{fig:LambdaIM}
\end{figure}

Similar to the extraction of anisotropic flow in heavy-ion collisions, the invariant--mass (IM) method~\cite{STAR:2018gyt} can be employed to extract the global polarization signal, where the value of $\langle \sin(\Psi_{1}-\phi_D^{\ast})\rangle$ in Eq.~(\ref{LambdaPHfunction}) is measured as a function of the invariant mass, $M_{\rm inv}$, as shown in Fig.~\ref{fig:LambdaIM}.
The observed polarization correlator $\langle \sin(\Psi_{1}-\phi_D^{\ast})\rangle^{\rm obs}$ contains contributions from both signal and background, denoted as ``$\mathrm{Sg}$'' and ``$\mathrm{Bg}$'', respectively:
\begin{equation}
\begin{split}
\langle \sin(\Psi_{1}-\phi_D^{*}) \rangle^{\rm obs}  =  (1-f^{\rm Bg}(M_{\rm inv})) \langle\sin(\Psi_{1}-\phi_D^{\ast})\rangle^{\rm Sg} \\
~+~f^{\rm Bg}(M_{\rm inv})\langle\sin(\Psi_{1}-\phi_D^{\ast})\rangle^{\rm Bg},
\end{split}\label{eq:IMmethod}
\end{equation} 
where $f^{\rm Bg}(M_{\rm inv})$ is the background fraction as a function of $M_{\rm inv}$, which is extracted by fitting the invariant mass distribution. 
The background contribution $\langle\sin(\Psi_{1}-\phi_D^{\ast})\rangle^{\rm Bg}$ can arise due to misidentification of particles.
The data were fitted with Eq.~(\ref{eq:IMmethod}) to extract the polarization signal.
The background contribution is statistically consistent with zero and is considered a source of systematic uncertainty (see Sec.~\ref{syst_section}). 

For $\Omega^{-}$ hyperons, the IM method is insufficient for reliable polarization extraction due to the large background fraction in the $\Omega^{-}$ invariant mass distribution. In this method, the background contribution to the polarization correlator $\langle\sin(\Psi_1 - \phi_D^*)\rangle$ is typically assumed to be zero or described by a smooth function. For $\Omega^{-}$ hyperons, however, this assumption is not justified: the combinatorial background contains real $\Lambda$ hyperons from other processes, which themselves carry a physical polarization signal. This introduces a nonzero effective background polarization that cannot be subtracted using the conventional IM approach.
To overcome this limitation, we have developed a two-stage extraction procedure. In the first stage, the polarization of daughter $\Lambda$ hyperons is extracted as a function of invariant mass using the standard IM method. In the second stage, the $\Omega^{-}$ mass region is divided into several bins; for each bin, the daughter $\Lambda$ polarization is extracted. 
The $\Omega^{-}$ polarization is then extracted from the mass dependence of the daughter-$\Lambda$ polarization accounting for the spin-transfer factor $C_{\Omega\Lambda}$, with the background contribution constrained by the mass-sideband regions. 
This procedure effectively separates the genuine $\Omega^{-}$ signal from background contributions.
As a consistency check of the two-stage procedure, the same method is applied to the $\Xi$ sample, for which the signal-to-background ratio is substantially higher. 
The polarization obtained with the two-stage method is consistent with that from the conventional invariant-mass approach within uncertainties.

When the IM method is applied separately to the positive- and negative-$\eta$ regions, the measured polarization correlator exhibits an apparent asymmetry, primarily because the directed-flow contribution ($v_1$) changes sign across midrapidity. 
For a symmetric $\eta$ interval, such as $-1<\eta<1$, the $v_1$ contributions largely cancel, yielding an unbiased $\eta$-integrated polarization. To determine the rapidity dependence, however, the IM extraction must be supplemented by a $v_1$ fit to remove the directed-flow contribution and isolate the genuine polarization signal. Here, $\phi_p^*$ denotes the azimuthal angle of the daughter proton in the $\Lambda$ rest frame and is therefore the relevant $\phi_D^*$ for this measurement.
In practice, the $v_1$ contribution is constrained by fitting the observed correlator $\langle \sin(\Psi_1 - \phi_p^*) \rangle$ together with the relative-angle correlator $\langle\sin(\phi_\Lambda - \phi_p^*)\rangle$, allowing extraction of the global polarization after removing $v_1$ contributions:
\begin{align}
    \langle \sin(\Psi_1 - \phi_p^*) \rangle^{\rm obs} = \langle \sin(\Psi_1 - \phi_p^*) \rangle^{\rm true} + c v_1 \langle \sin(\phi_\Lambda - \phi_p^*) \rangle.
\end{align}
Here, the superscripts ``${\rm obs}$'' and ``${\rm true}$'' denote the measured and $v_1$-corrected polarization correlators, respectively, and $v_1$ denotes the directed-flow coefficient. 
Further details on this procedure can be found in Ref.~\cite{STAR:2021beb}.

\subsection{\label{Systematic uncertainties}Systematic uncertainties}
\label{syst_section}

The main sources of systematic uncertainty include uncertainties in the hyperon decay parameters, background contributions, the variation of KFParticle reconstruction parameters, and the event-plane determination.
Contributions from the event-plane resolution and the detector acceptance are found to be small ($<0.1\%$).

The uncertainties associated with the hyperon decay parameters arise from the limited precision of the known weak decay asymmetries. For $\Lambda$ and $\bar{\Lambda}$, we take $\alpha_{\Lambda} = -\alpha_{\bar{\Lambda}} = 0.732 \pm 0.014$, corresponding to a relative uncertainty of 1.9\%. For $\Xi^{-}$ and $\bar{\Xi}^{+}$, $\alpha_{\Xi^{-}} = -\alpha_{\bar{\Xi}^{+}} = -0.401 \pm 0.010$, giving a relative uncertainty of 2.5\%. For $\Omega^{-}$ hyperons, the polarization is measured indirectly via the daughter $\Lambda$, resulting in a relative uncertainty of 1.9\%.

The systematic uncertainty associated with background extraction is evaluated by replacing the default second-order polynomial with a first-order polynomial. The representative relative uncertainties are approximately 2\% for $\Lambda$, 8\% for $\Xi$, and 30\% for $\Omega$. For $\Lambda$ and $\Xi$, the uncertainty is evaluated separately for each collision energy and centrality interval; the 30\% relative uncertainty for $\Omega$ is applied to all reported measurements. The larger $\Omega$ uncertainty reflects its lower signal-to-background ratio and the sensitivity of the extraction to the mass-dependent polarization of the residual background.

The influence of the KFParticle reconstruction parameters, including the decay-length and DCA selections, is evaluated by varying the corresponding cuts within the prescribed ranges. For each source, the associated systematic uncertainty is assigned as the maximum absolute deviation of the polarization from the default result among these variations. 
The resulting relative uncertainties, applied uniformly across all collision energies and centrality intervals, are approximately 3\% for the direct $\Xi$ measurement, 4\% for the indirect $\Xi$ measurement, and 6\% for the $\Omega^{-}$ measurement.

The default event plane is determined from the correlation between the $\Psi_1$ measurements from the two EPD wheels at forward and backward rapidities. The systematic uncertainty associated with this determination is estimated by comparing $\Psi_1$ reconstructed using a single EPD side (East or West) to that using the full EPD.
For each collision energy and centrality bin, the relative differences $|P_H^{\text{East}}-P_H^{\text{Full}}|$ and $|P_H^{\text{West}}-P_H^{\text{Full}}|$ are computed, and their average is taken as the systematic uncertainty for that bin. 
These uncertainties are averaged within the 0--20\%, 20--50\%, and 50--80\% centrality intervals, and the resulting value is applied to all measurements in the corresponding interval at that energy.
The same procedure is applied to $\Lambda$, $\Xi$, and $\Omega^{-}$, as they use the same $\Psi_1$. The representative relative systematic uncertainties, averaged over the BES-II energies within each centrality interval, are $\sim$15.6\%, $\sim$3.5\%, and $\sim$4.0\%, respectively.


For each measurement, the contributions from different systematic
sources are treated as independent and combined in quadrature
to obtain the total systematic uncertainty.
Representative values for the systematic uncertainties are summarized in Tables~\ref{tab:LambdaSys}, \ref{tab:XiDirectSys}, \ref{tab:XiIndirectSys}, and \ref{tab:OmegaSys}.


\begin{table}[h]
\centering
\caption{Summary of representative systematic uncertainties for $\Lambda$ hyperons.}
\begin{tabular}{l c}
\hline
Source & Relative uncertainty (\%) \\
\hline
Decay parameter           & \phantom{0}1.9 \\
Background subtraction    & 1.5--3.5 \\
Event-plane determination & 3.5--15.6 \\
Event-plane resolution    & $<0.1$ \\
Acceptance                & $<0.1$ \\
\hline
Total                     & 4.3--16.1 \\
\hline
\end{tabular}
\label{tab:LambdaSys}
\end{table}

\begin{table}[h]
\centering
\caption{Summary of representative systematic uncertainties for $\Xi$ hyperons from direct measurement. }
\begin{tabular}{l c}
\hline
Source & Relative uncertainty (\%) \\
\hline
Decay parameter & 2.5 \\
Background subtraction & 7--12 \\
KFParticle variations & 3 \\
Event-plane determination & 3.5--15.6\\
Event-plane resolution & $<0.1$ \\
Acceptance & $<0.1$ \\
\hline
Total & 8.7--20.1 \\
\hline
\end{tabular}
\label{tab:XiDirectSys}
\end{table}

\begin{table}[h]
\centering
\caption{Summary of representative systematic uncertainties for $\Xi$ hyperons from indirect measurement. }
\begin{tabular}{l c}
\hline
Source & Relative uncertainty (\%) \\
\hline
Decay parameter & 1.9 \\
Background subtraction & 3--6 \\
KFParticle variations & 4 \\
Event-plane determination & 3.5--15.6\\
Event-plane resolution & $<0.1$ \\
Acceptance & $<0.1$ \\
\hline
Total & 6.4--17.3 \\
\hline
\end{tabular}
\label{tab:XiIndirectSys}
\end{table}

\begin{table}[h]
\centering
\caption{Summary of representative systematic uncertainties for $\Omega^{-}$ hyperons.}
\begin{tabular}{l c}
\hline
Source & Relative uncertainty (\%) \\
\hline
Decay parameter & 1.9 \\
Background subtraction & 30 \\
KFParticle variations & 6 \\
Event-plane determination & 3.5--15.6\\
Event-plane resolution & $<0.1$ \\
Acceptance & $<0.1$ \\
\hline
Total & 30.9--34.4 \\
\hline
\end{tabular}
\label{tab:OmegaSys}
\end{table}

\section{\label{sec:results}results}

\subsection{Centrality dependence of global polarization}
\label{sec:Centresults}

\begin{figure*}[t]
    \centering
    \includegraphics[width=0.88\linewidth]{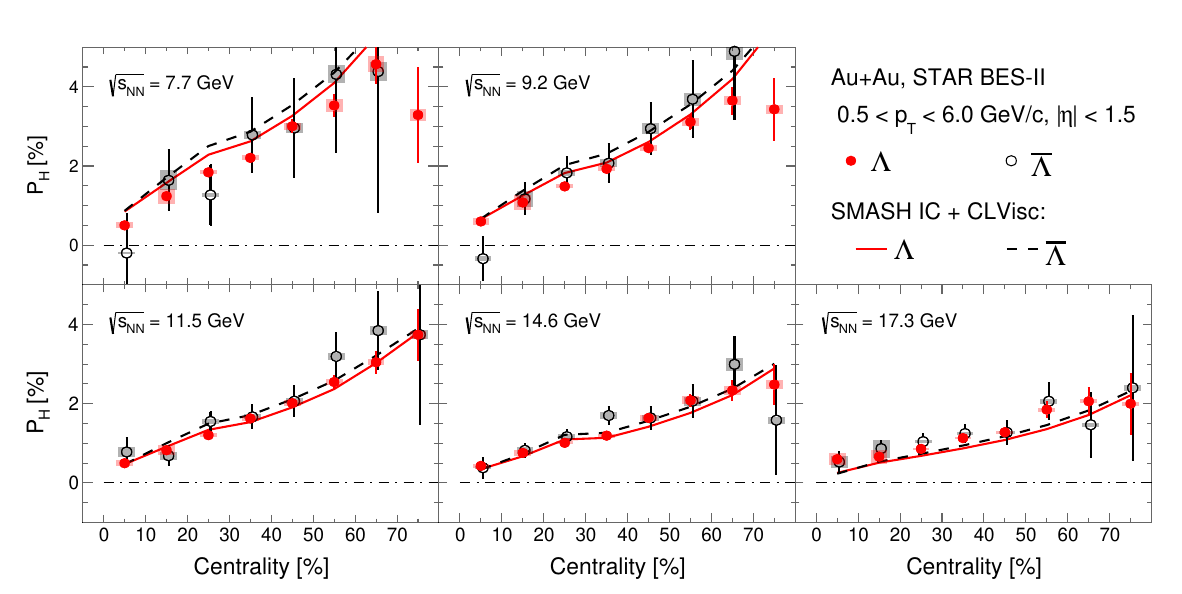}
    \caption{Global polarization of $\Lambda$ and $\bar{\Lambda}$ hyperons as a function of collision centrality at $\sqrt{s_\mathrm{NN}} = 7.7, 9.2, 11.5, 14.6,$ and $17.3$ GeV. The experimental data are inclusive and are not corrected for feed-down. The polarization results are compared with SMASH IC+CLVisc model calculations. Calculations for primary $\Lambda$ and $\bar{\Lambda}$ are shown as solid and dashed lines, respectively~\cite{Li:2026udm}. The statistical uncertainties are shown as error bars, while the systematic uncertainties are shown as open boxes.}
    \label{fig:LamCentDep}
\end{figure*}

\begin{figure*}[t]
    \centering
    \includegraphics[width=0.95\linewidth]{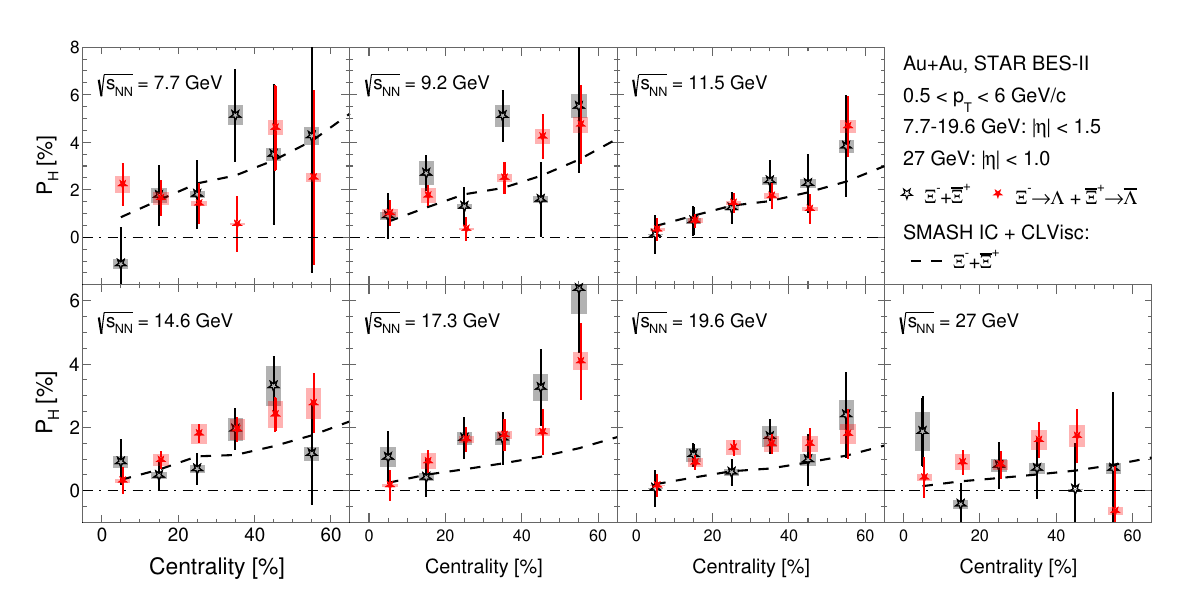}
    \caption{Global polarization of $\Xi^{-}+\bar{\Xi}^{+}$ hyperons as a function of collision centrality at $\sqrt{s_\mathrm{NN}} = 7.7, 9.2, 11.5, 14.6, 17.3, 19.6$, and 27 GeV. The experimental data are not corrected for feed-down. The data are compared with SMASH IC+CLVisc model calculations for primary $\Xi^{-} + \bar{\Xi}^{+}$~\cite{Li:2026udm}. The statistical uncertainties are shown as error bars, while the systematic uncertainties are shown as open boxes.}
    \label{fig:XiCentDep}
\end{figure*}

\begin{figure*}[t]
    \centering
    \includegraphics[width=0.95\linewidth]{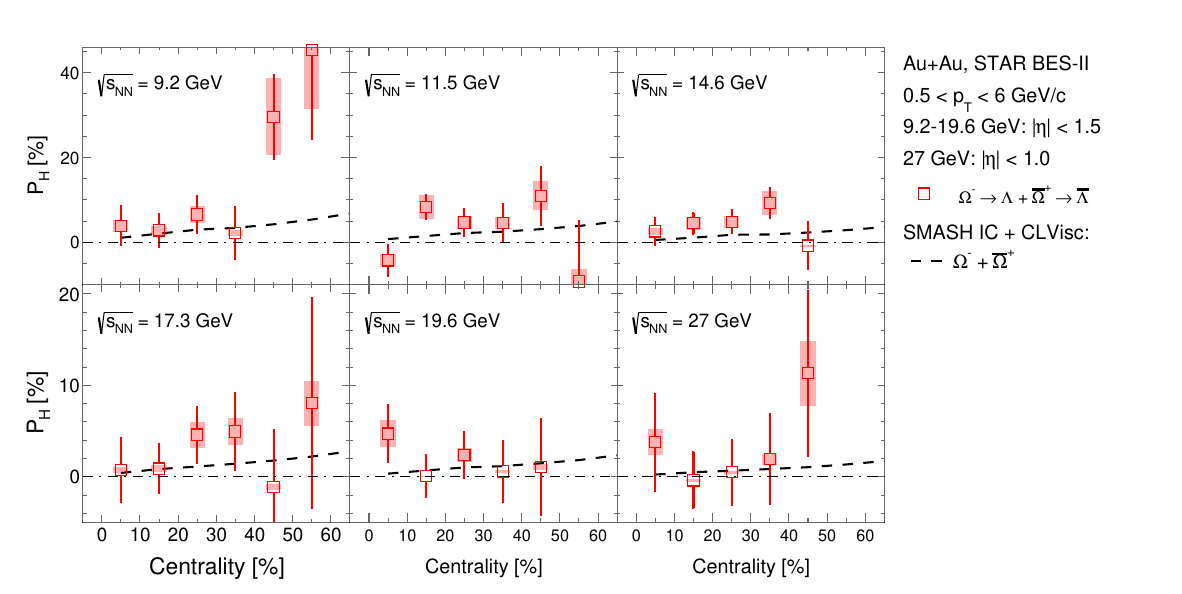}
    \caption{Global polarization of $\Omega^{-}+\bar{\Omega}^{+}$ hyperons as a function of collision centrality at $\sqrt{s_\mathrm{NN}} = 9.2, 11.5, 14.6, 17.3, 19.6$, and 27 GeV. The polarization results are compared with SMASH IC+CLVisc model calculations. The dashed lines show model results for $\Omega^{-} + \bar{\Omega}^{+}$~\cite{Li:2026udm}. The statistical uncertainties are shown as error bars, while the systematic uncertainties are shown as open boxes.}
    \label{fig:OmegaCentDep}
\end{figure*}

We present the centrality dependence of the global polarization of $\Lambda$, $\Xi$, and $\Omega$ hyperons in Au+Au collisions in Figs.~\ref{fig:LamCentDep}--\ref{fig:OmegaCentDep}. 
For $\Lambda$ and $\bar{\Lambda}$ hyperons, Fig.~\ref{fig:LamCentDep} shows a clear global polarization signal that increases toward peripheral collisions. 
Figure~\ref{fig:XiCentDep} presents the centrality dependence of the $\Xi^{-}+\bar{\Xi}^{+}$ global polarization obtained with the two methods defined in Sec.~\ref{sec:PHextract}. 
The direct and indirect measurements both exhibit a centrality dependence and agree within uncertainties. 
The direct $\Xi$ measurement has larger uncertainties than that inferred from the daughter $\Lambda$ because the decay parameter $|\alpha_{\Xi}| = 0.401$ is roughly half of $|\alpha_{\Lambda}| = 0.732$. 
Within uncertainties, the centrality dependence of the $\Xi^{-}+\bar{\Xi}^{+}$ polarization is consistent with that of the inclusive $\Lambda+\bar{\Lambda}$ polarization.
Figure~\ref{fig:OmegaCentDep} presents the corresponding centrality dependence for $\Omega$ hyperons, with statistical uncertainties larger than those for $\Lambda$ and $\Xi$ due to the lower $\Omega$ production yield.
The reported $\Omega$ polarization is inferred from the daughter-$\Lambda$ polarization using $C_{\Omega\Lambda}\approx+1$, as described in Sec.~\ref{sec:PHextract}.
Under the common global-polarization interpretation, this choice corresponds to an inferred $\Omega$ polarization with the same orientation as the measured $\Lambda$ and $\Xi$ polarizations.
Overall, the measured global polarizations of all three hyperon species generally increase toward more peripheral collisions, and no significant difference in the centrality dependence among $\Lambda$, $\Xi$, and $\Omega$ is resolved within the current uncertainties.

For comparison with the differential measurements, we use the SMASH~IC+CLVisc calculation~\cite{Li:2026udm}, which combines SMASH initial conditions with $(3+1)$-dimensional viscous hydrodynamic evolution.
Hyperon polarization is evaluated using a modified Cooper--Frye formalism including contributions from thermal vorticity, thermal shear, and the gradient of the baryon chemical potential~\cite{Li:2026udm}.
For the differential observables considered here, the calculations correspond to primary hyperons and do not include feed-down contributions from decays of heavier hadrons~\cite{Li:2026udm}, whereas the experimental measurements are inclusive.
The calculations qualitatively reproduce the increase in polarization toward more peripheral collisions for $\Lambda$, $\Xi$, and $\Omega$.
Within this framework, the stronger velocity gradients in more peripheral collisions generate larger vorticity, leading to an enhanced global polarization~\cite{Li:2026udm}.
The calculations further predict $P_{\Xi}\gtrsim P_{\Lambda}$ and $P_{\Omega}\simeq(5/3)P_{\Lambda}$ at fixed centrality~\cite{Li:2026udm}.
Within the current experimental uncertainties, however, no significant species ordering is resolved in the data.
Independent model calculations indicate that feed-down reduces the inclusive $\Lambda$ polarization by about 10\% and increases the inclusive $\Xi^{-}$ polarization by about 25\% relative to their primary values~\cite{Li:2021zwq}. 
This opposite effect results from the different decay contributions to the inclusive $\Lambda$ and $\Xi^{-}$ samples, through which the polarizations of heavier parent particles are transferred to the daughter hyperons~\cite{Li:2021zwq}.
The model--data comparison is therefore restricted to a qualitative assessment.

\subsection{Transverse momentum ($p_{T}$) dependence of global polarization}

\begin{figure*}[t]
    \centering
    \includegraphics[width=0.87\linewidth]{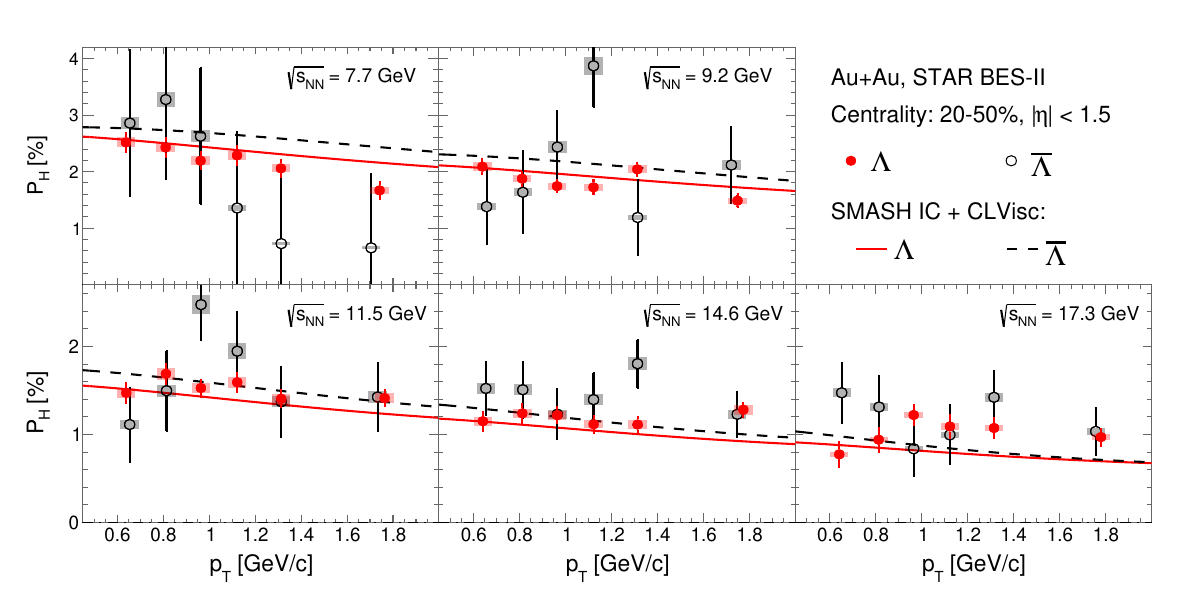}
    \caption{Global polarization of $\Lambda$ and $\bar{\Lambda}$ as a function of $p_T$ in Au+Au collisions from BES-II. The polarization results are compared with SMASH IC+CLVisc model calculations. Calculations for primary $\Lambda$ and $\bar{\Lambda}$ are shown as solid and dashed lines, respectively~\cite{Li:2026udm}.}
    \label{fig:LamPtDep}
\end{figure*}

\begin{figure*}[t]
    \centering
    \includegraphics[width=0.95\linewidth]{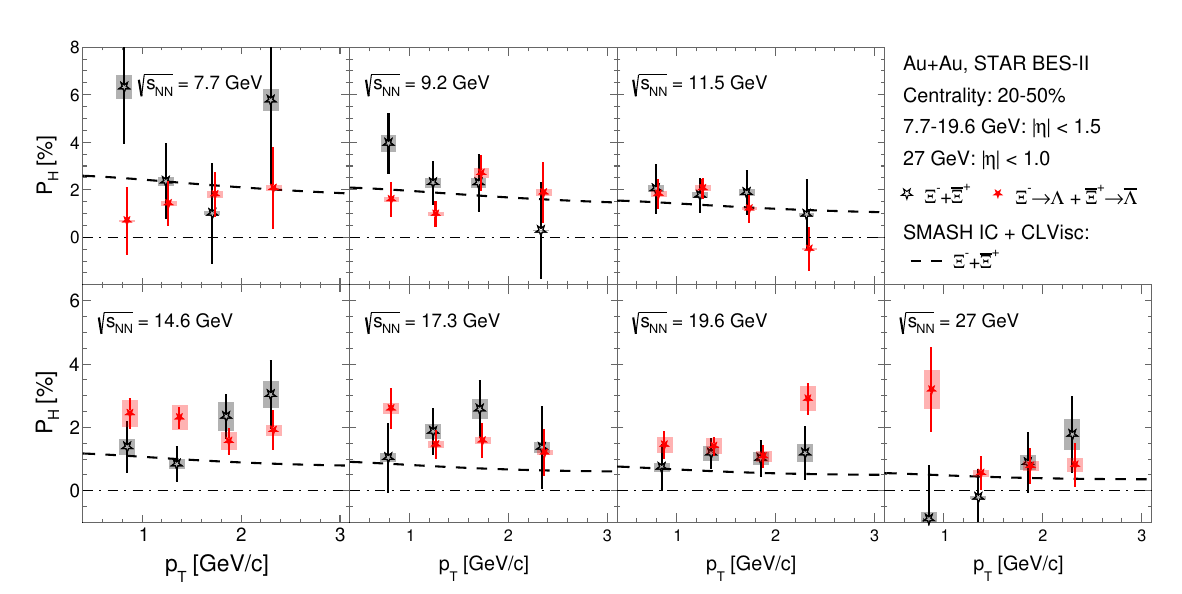}
    \caption{Global polarization of $\Xi^{-} + \bar{\Xi}^{+}$ as a function of $p_T$ in Au+Au collisions from BES-II. The polarization results are compared with SMASH IC+CLVisc model calculations. The dashed lines show model results for primary $\Xi^{-} + \bar{\Xi}^{+}$~\cite{Li:2026udm}.}
    \label{fig:XiPtDep}
\end{figure*}

We have also studied the dependence of the global polarization on transverse momentum, $p_T$, for $\Lambda$ and $\Xi$ hyperons. Figure~\ref{fig:LamPtDep} shows the polarization of $\Lambda$ and $\bar{\Lambda}$ as a function of $p_T$ in the 20--50\% centrality range. Within the uncertainties, no significant $p_T$ dependence is observed. Similar behavior is seen for $\Xi$ hyperons in Fig.~\ref{fig:XiPtDep}, where the polarization remains approximately constant across the measured $p_T$ range. 
No results on the $p_T$ dependence are shown for $\Omega$
hyperons due to limited statistics.

Theoretically, hyperon polarization may acquire a $p_T$ dependence from the spatial structure of thermal vorticity and other dynamical effects~\cite{Wei:2018zfb,Guo:2021udq,Becattini:2016gvu}. The observed flat dependence disfavors a strong momentum variation over the measured range and is compatible with SMASH~IC+CLVisc calculations~\cite{Li:2026udm}, which predict only a weak dependence for primary $\Lambda$ and $\Xi$ hyperons. The current precision does not discriminate among weakly varying scenarios.

\subsection{Rapidity dependence of global polarization}

\begin{figure*}[t]
    \centering
    \includegraphics[width=0.95\linewidth]{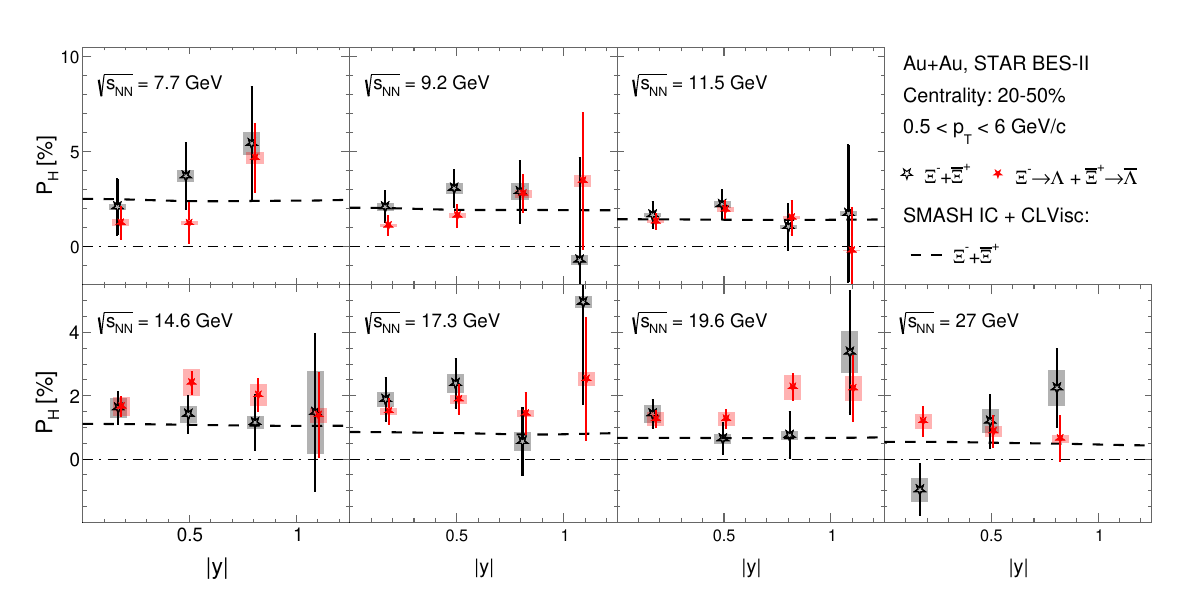}
    \caption{Global polarization of $\Xi^{-} + \bar{\Xi}^{+}$ as a function of $|y|$ in Au+Au collisions from BES-II. The polarization results are compared with SMASH IC+CLVisc model calculations. The dashed lines show model results for primary $\Xi^{-} + \bar{\Xi}^{+}$~\cite{Li:2026udm}.}
    \label{fig:XiRapDep}
\end{figure*}

Figure~\ref{fig:XiRapDep} shows the $\Xi$ polarization as a function of the absolute value of the hyperon rapidity, $|y|$.
To quantify its rapidity dependence, the measurements at
$\sqrt{s_{\mathrm{NN}}}=7.7$--27~GeV are combined within each $|y|$
interval and fitted with the linear function $P=a+b|y|$.
The resulting slopes are $b=0.07\pm0.70$ for the direct measurement~($\Xi^{-}+\bar{\Xi}^{+}$) and
$b=0.71\pm0.44$ for the indirect measurement~($\Xi^{-}\rightarrow\Lambda+\bar{\Xi}^{+}\rightarrow\bar{\Lambda}$),
where the uncertainties include both statistical and systematic contributions.
Both slopes are consistent with zero within uncertainties, and no statistically
significant rapidity dependence of the $\Xi$ polarization is resolved at the
present precision. The direct method has substantially larger statistical
uncertainties than the indirect method because of the smaller weak-decay
parameter $|\alpha_{\Xi}|$, while the results obtained with the two methods
are mutually consistent.
For comparison, the combined $\Lambda$ measurement over
$\sqrt{s_{\mathrm{NN}}}=7.7$--17.3~GeV yields a slope of
$b=0.498\pm0.103$, which differs from zero by $4.8$ standard deviations~\cite{shortpaper}.
Although a significant rapidity dependence is therefore resolved for $\Lambda$
but not for $\Xi$, the fitted slopes are statistically compatible within the
current $\Xi$ uncertainties. The present data thus do not establish a
species-dependent difference in the rapidity evolution.

The observed $\Lambda$ rapidity dependence is sensitive to the
three-dimensional deposition of angular momentum, longitudinal shear,
baryon stopping, and the space--time evolution of thermal vorticity~\cite{Liang:2019pst,Sun:2017xhx,Xie:2019jun,Li:2026udm}.
Whether different hyperon species sample this underlying vortical structure
differently through hadronization, feed-down, freeze-out, or subsequent
hadronic interactions remains an open question.
The SMASH~IC+CLVisc calculations predict a weak rapidity dependence for the primary hyperons. Within the present uncertainties, this behavior is compatible with the $\Xi$ measurements, whereas the measured $\Lambda$ polarization exhibits a stronger rapidity dependence, particularly at the lower collision energies~\cite{Li:2026udm}. 
Since the measurements are inclusive whereas the calculations presented here correspond to primary hyperons, a quantitative species comparison requires a consistent treatment of feed-down, experimental acceptance, and species-dependent freeze-out. These effects may also modify the rapidity dependence of the inclusive polarization; however, the present data do not allow them to be separated experimentally.
Calculations including feed-down and the experimental acceptance will be required to determine whether this difference reflects species-dependent dynamics or the current statistical sensitivity, while higher-precision multistrange measurements would provide a further test.



\section{\label{sec:summary}Summary}

We have reported high-precision measurements of the global polarization of $\Lambda$, $\Xi$, and $\Omega$ hyperons in Au+Au collisions at $\sqrt{s_{\mathrm{NN}}}=7.7$--$27~\mathrm{GeV}$ from the RHIC BES-II program using the STAR detector. The upgraded inner TPC and Event Plane Detector enable improved tracking, particle identification, and first-order event-plane resolution, facilitating precise measurements of hyperon polarization across a wide kinematic range.
The measurements exhibit a clear centrality dependence, with larger polarization values in more peripheral collisions, qualitatively consistent with vorticity-based calculations of the centrality dependence of global polarization.
No significant dependence of polarization on transverse momentum is observed for $\Lambda$ and $\Xi$. The $\Lambda$ polarization reveals a rapidity dependence that is more prominent at lower collision energies, while no significant rapidity dependence is observed for $\Xi$ within current precision.
Within current precision and under the stated spin-transfer assumptions, the $\Xi$ and $\Omega$ measurements are consistent with the $\Lambda$ results and with a common vorticity-driven polarization mechanism across strange and multistrange hyperons. Together with the centrality, $p_T$, and rapidity dependences, these results provide a coherent differential test of vorticity-driven polarization. The data leave room for additional contributions from flavor-dependent dynamics, feed-down, species-dependent freeze-out, and hadronic interactions, but higher-statistics $\Omega$ measurements and improved theoretical calculations are required to determine whether a genuine species hierarchy exists.

\section*{Acknowledgements}
We thank the RHIC Operations Group and SCDF at BNL, the NERSC Center at LBNL, and the Open Science Grid consortium for providing resources and support.  This work was supported in part by the Office of Nuclear Physics within the U.S. DOE Office of Science, the U.S. National Science Foundation, National Natural Science Foundation of China, Chinese Academy of Science, the Ministry of Science and Technology of China and the Chinese Ministry of Education, NSTC Taipei, the National Research Foundation of Korea, Czech Science Foundation and Ministry of Education, Youth and Sports of the Czech Republic, Hungarian National Research, Development and Innovation Office, New National Excellency Programme of the Hungarian Ministry of Human Capacities, Department of Atomic Energy and Department of Science and Technology of the Government of India, the National Science Centre and WUT ID-UB of Poland, German Bundesministerium f\''ur Bildung, Wissenschaft, Forschung and Technologie (BMBF), Helmholtz Association, Ministry of Education, Culture, Sports, Science, and Technology (MEXT), Japan Society for the Promotion of Science (JSPS), and Agencia Nacional de Investigacion y Desarrollo de Chile (ANID), Chile.

\bibliography{ref}

\end{document}